\documentclass[reprint,nofootinbib,aps,superscriptaddress,amsmath,amssymb,onecolumn,11pt]{revtex4-2}

\usepackage{graphicx}
\usepackage{dcolumn}
\usepackage{bm}
\usepackage{braket}
\usepackage{breqn}
\usepackage{mathrsfs}
\usepackage{color}

\usepackage[colorlinks=true,linkcolor=red,citecolor=blue]{hyperref}

\begin{document}

\title{Charged Superradiant Instability of Spherically Symmetric Regular Black Holes in de Sitter Spacetime: Time- and Frequency-Domain Analysis}

\author{Yizhi Zhan}
\email{niannian\_12138@163.com}

\author{Hengyu Xu}
\email{xuhengyu0501@outlook.com}

\author{Haowei Chen}
\email{chenhaowei@zjut.edu.cn}

\author{Shao-Jun Zhang}
\email{sjzhang@zjut.edu.cn (corresponding author)}

\affiliation{Institute for Theoretical Physics and Cosmology$,$ Zhejiang University of Technology$,$ Hangzhou 310032$,$ China}
\affiliation{School of Physics$,$ Zhejiang University of Technology$,$ Hangzhou 310032$,$ China}
\date{\today}

\begin{abstract}

We investigate the superradiant instability of Ay\'on-Beato-Garc\'ia-de Sitter (ABG-dS) black holes under massless charged scalar perturbations using both time-domain evolutions and frequency-domain computations. We show that the instability occurs only for the spherically symmetric mode with $\ell=0$, whereas asymptotically flat ABG black holes remain stable in the massless limit, which underscores the essential role of the cosmological horizon in providing a confining boundary. We further study the dependence of the growth rate on the cosmological constant $\Lambda$, the scalar charge $q$, and the black hole charge $Q$, finding that it reaches a maximum at intermediate values of $\Lambda$ and $q$ and increases monotonically with $Q$. Compared with Reissner-Nordstr\"om-de Sitter black holes, ABG-dS black holes exhibit distinct instability characteristics due to the modified electrostatic potential induced by nonlinear electrodynamics.

\end{abstract}

\maketitle
  
\section{Introduction}

Recent observational advances have made black holes a primary arena for testing gravity in the strong-field regime. Landmark examples include the first direct detections of gravitational waves \cite{LIGOScientific:2016aoc,LIGOScientific:2016sjg} and horizon-scale images of black hole shadows \cite{EventHorizonTelescope:2019dse,EventHorizonTelescope:2019ggy}. These developments motivate detailed theoretical investigations of black hole spacetimes and, in particular, their dynamical response to perturbations.

A long-standing conceptual challenge is the presence of spacetime singularities in many black hole solutions. In Einstein gravity, the singularity theorems of Penrose, Hawking, and others imply that, under standard classical assumptions, singularities generically form during gravitational collapse \cite{Penrose:1964wq,Hawking:1973uf,Senovilla:1998oua}. Since curvature invariants and energy densities diverge at a singularity, it is natural to interpret it as signaling a breakdown of the classical description and the need for new physics.

One approach to addressing this issue is to construct regular black holes whose geometry remains nonsingular everywhere. Such solutions typically require violations or effective modifications of the classical energy conditions, which may arise from quantum effects or from nonlinear matter sectors. A prototypical example is the regular model proposed by Bardeen \cite{1968qtr..conf...87B}. Ay\'on-Beato and Garc\'ia later provided an explicit realization within Einstein gravity coupled to nonlinear electrodynamics and obtained the family now known as ABG black holes \cite{Ayon-Beato:1998hmi,Ayon-Beato:1999kuh}, together with a nonlinear-electrodynamic interpretation of the Bardeen solution \cite{Ayon-Beato:2000mjt}. Other representative regular models include the Hayward and Dymnikova solutions, as well as regular black holes sourced by nonlinear electrodynamics \cite{Hayward:2005gi,Dymnikova:1992ux,Bronnikov:2000vy,Dymnikova:2004zc}. Systematic constructions and recent overviews can be found in \cite{Fan:2016hvf,Lan:2023cvz,Torres:2022twv,Bronnikov:2022ofk,Bueno:2024dgm,Ovalle:2023ref}; these models can be viewed as effective descriptions of short-distance corrections to Maxwell theory \cite{Heisenberg:1936nmg,Born:1934gh,Polchinski:1998rq,Polchinski:1998rr}.

A second ingredient of direct relevance to realistic astrophysical and cosmological settings is a positive cosmological constant. Observations of distant supernovae \cite{SupernovaSearchTeam:1998fmf,SupernovaCosmologyProject:1998vns} and the cosmic microwave background \cite{Planck:2018vyg} support an accelerating universe, which is well modeled by $\Lambda>0$. Black holes in de Sitter backgrounds possess both an event horizon and a cosmological horizon, which yields a richer causal structure and can lead to dynamical phenomena absent in asymptotically flat spacetimes \cite{Gibbons:1977mu,Brady:1996za,Brady:1999wd,Molina:2003dc}.

The stability of de Sitter black holes has therefore been studied extensively. Schwarzschild-de Sitter black holes are stable against neutral scalar, electromagnetic, and gravitational perturbations \cite{Brady:1996za,Brady:1999wd,Molina:2003dc,Cardoso:2003vt}. For Reissner-Nordstr\"om de Sitter (RN-dS) black holes, charged scalar perturbations can trigger an instability that appears only for the spherically symmetric mode with $\ell=0$ \cite{Zhu:2014sya}. The underlying mechanism combines superradiant amplification with the trapping of the amplified field in an effective potential well, and it can be quenched by a sufficiently large scalar mass \cite{Konoplya:2014lha}.

Superradiance and its associated instabilities have a long history and arise whenever superradiantly amplified modes are confined. While the classic and most prominent examples involve rotating black holes \cite{Damour:1976kh,Detweiler:1980uk,Zouros:1979iw,Cardoso:2005vk,Dolan:2007mj,Konoplya:2013rxa}, charged superradiance can also occur in spherically symmetric spacetimes. In the asymptotically flat Reissner-Nordstr\"om case, however, superradiance alone does not typically lead to an instability unless an additional confinement mechanism is present \cite{Hod:2012wmy,Hod:2013nn}. Such confinement can be provided either by boundary conditions or by the spacetime structure itself \cite{Herdeiro:2013pia,Zhu:2014sya,Konoplya:2014lha,Dolan:2015dha,Dias:2018zjg,Davey:2021oye,Richarte:2021fbi,Feiteira:2024awb}. Recent works suggest that nonlinear electromagnetic effects in regular black holes may open new channels for superradiant instabilities even in spherical symmetry \cite{Hod:2024aen,dePaula:2024xnd,Dolan:2024qqr}.

Motivated by these developments, we study the superradiant instability of ABG black holes in a de Sitter background under massless charged scalar perturbations. The cosmological horizon provides a natural outer boundary that can confine superradiant modes and thereby qualitatively modify the stability properties. We show that, although asymptotically flat ABG black holes are stable in the massless limit, ABG-dS black holes can become unstable even for massless fields. We employ both time-domain evolutions and frequency-domain calculations to identify unstable modes and to quantify the instability across representative ranges of the black hole charge $Q$, the scalar charge $q$, and the cosmological constant $\Lambda$. We further compare our results with the RN-dS case and highlight the role of the modified electrostatic potential associated with nonlinear electrodynamics.

This paper is organized as follows. Section II reviews the ABG-dS spacetime. Section III derives the equation of motion for a charged scalar field perturbation. Sections IV and V present numerical results from time domain evolution and from frequency domain analysis, respectively. The final section summarizes our conclusions and comments on possible extensions. We work in natural units with $G=c=\hbar=4\pi\epsilon_0=1$, where $G$ is Newton's gravitational constant, $c$ is the speed of light in vacuum, $\hbar$ is the reduced Planck constant, and $\epsilon_0$ is the vacuum permittivity.

\section{Ayón-Beato-García-de Sitter black hole}

The metric of the Ayón-Beato-García-de Sitter (ABG-dS) black hole can be written as \cite{Mo:2006tb}
    \begin{align}
    ds^2 &=-f(r)dt^2+f(r)^{-1}dr^2+r^2 (d\theta^2 + \sin^2 \theta d \phi^2),\\
    f(r) &=1-\frac{2Mr^2}{(r^2+Q^2)^{3/2}}+\frac{Q^2 r^2}{(r^2+Q^2)^2}-\frac{\Lambda r^2}{3},
\end{align}
where $M$ and $Q$ denote the mass and charge parameters of the black hole, and $\Lambda$ is the cosmological constant. The electromagnetic vector potential is taken as $A_\mu=(-\Phi(r),0,0,0)$, where
    \begin{align}
    \Phi(r)=\frac{r^5}{2Q}\left(\frac{3M}{r^5}+\frac{2Q^2}{(Q^2+r^2)^3}-\frac{3M}{(Q^2+r^2)^{5/2}}\right).
\end{align}
This spacetime is a spherically symmetric, electrically charged, and regular solution of Einstein gravity coupled to nonlinear electrodynamics with a cosmological constant. In the limit $\Lambda \to 0$, the metric reduces to the well-known ABG black hole \cite{Ayon-Beato:1998hmi, Ayon-Beato:1999kuh}. The regularity of the spacetime at $r=0$ is reflected in the finiteness of curvature invariants, for example,
\begin{widetext}
    \begin{align}
    R & \approx 4\left(-\frac{3}{Q^2}+\frac{6M}{Q^3}+\Lambda\right)+\frac{(-90M+60Q)r^2}{Q^5}, 
    && r\rightarrow 0, \\
    R_{\mu\nu} R^{\mu\nu} &\approx \frac{4(6M-3Q+Q^3\Lambda)^2}{Q^6}-\frac{60 \left[ (3M-2Q)(6M-3Q+Q^3\Lambda) \right]r^2}{Q^8},
    && r\rightarrow 0, \\
    R_{\mu\nu\rho\sigma} R^{\mu\nu\rho\sigma} &\approx \frac{8(6M-3Q+Q^3\Lambda)^2}{3Q^6}-\frac{40 \left[ (3M-2Q)(6M-3Q+Q^3\Lambda) \right]r^2}{Q^8},
    && r\rightarrow 0.
\end{align}
\end{widetext}

\begin{figure}[!htbp]
    \centering
    \includegraphics[width=0.46\linewidth]{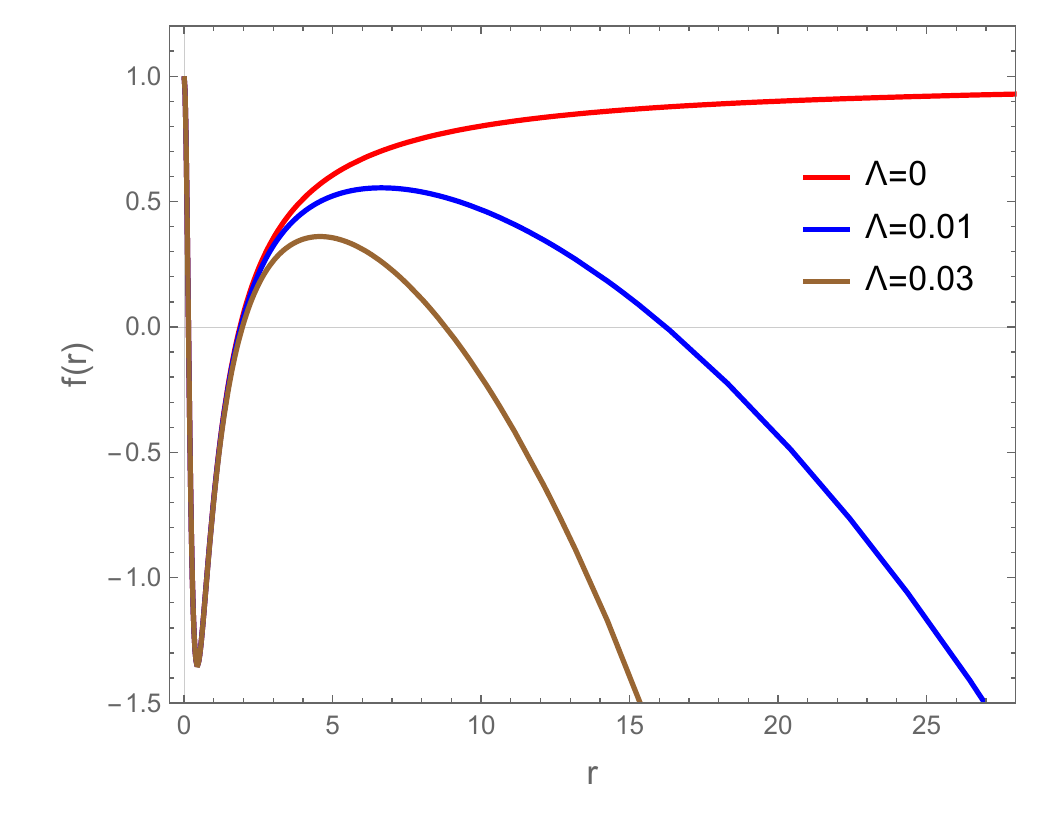}\quad
    \includegraphics[width=0.46\linewidth]{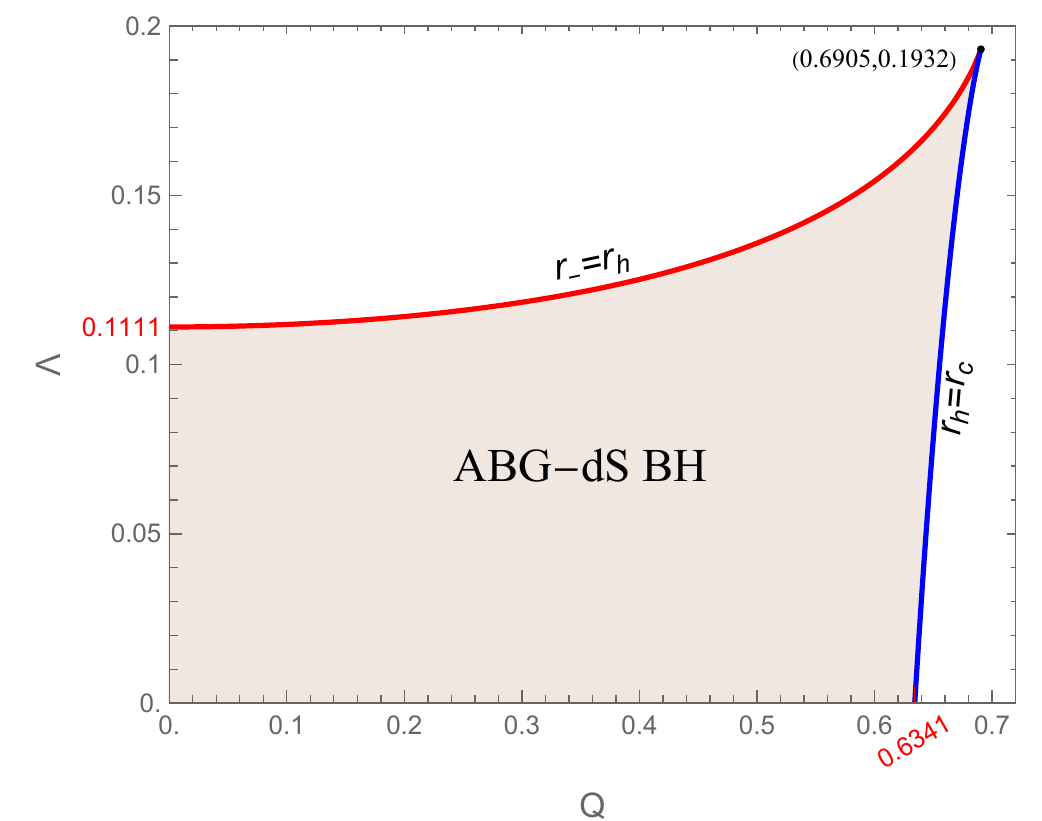}
    \caption{{\em Left}: Profile of the metric function $f(r)$ for various $\Lambda$, with $Q=0.3$. {\em Right}: Allowed region in the $Q-\Lambda$ plane for which the metric describes a black hole. The two critical curves $r_-=r_h$ and $r_h = r_c$ correspond to the two extremal limits.}
    \label{dSf}
\end{figure}

Throughout this work, we set $M=1$ and measure all physical quantities in units of the black hole mass. The left panel of Fig.~\ref{dSf} shows typical profiles of the metric function $f(r)$, which can admit up to three real roots corresponding to the inner horizon ($r=r_-$), the event horizon ($r=r_h$), and the cosmological horizon ($r=r_c$). The location of the cosmological horizon is particularly sensitive to $\Lambda$, and increasing $\Lambda$ decreases $r_c$. For the spacetime to describe a black hole, the horizon ordering $r_-\leq r_h\leq r_c$ must hold, which constrains the admissible region in the $Q$--$\Lambda$ parameter space, as shown in the right panel of Fig.~\ref{dSf}.

\section{Equation of charged scalar field perturbation}

We now consider a charged scalar field perturbation $\psi$ and investigate the possible onset of superradiant instability. Assuming that the scalar field is minimally coupled to the electromagnetic field, its dynamics is governed by the Klein--Gordon equation,
    \begin{align}
    (\nabla_\nu-iqA_\nu)(\nabla^\nu-iqA^\nu)\psi-\mu^2 \psi=0, \label{KGeq}
\end{align}
where $\mu$ and $q$ are the mass and charge of the field, respectively. For the time-domain analysis, we decompose the scalar field in spherical harmonics,
$\psi (t, r, \theta, \phi)= \sum_{\ell m}\frac{1}{r} \Psi_{\ell m}(t, r)Y_{\ell m}(\theta,\phi)$.
By spherical symmetry, modes with different $\ell$ decouple. For each fixed $\ell$, Eq.~(\ref{KGeq}) reduces to
\begin{align}
    -\frac{\partial^2 \Psi}{\partial t^2} + f(r)\frac{d}{dr}\left(f(r)\frac{d \Psi}{dr}\right)- 2 i q \Phi \frac{\partial \Psi}{\partial t} - V_I(r) \Psi = 0, \label{TimeRadialEq}
\end{align}
where the effective potential $V_I(r)$ is given by
    \begin{align}
    V_I(r)\equiv - q^2 \Phi^2 + f(r)\left[\mu^2+ \frac{f'(r)}{r}+\frac{\ell(\ell+1)}{r^2}\right], \label{EffectivePotential}
\end{align}
and we omit the subscripts from $\Psi_{\ell m}$ for simplicity.
 
For the frequency-domain analysis, we further decompose $\Psi(t, r) = \int d\omega \,u(r) e^{-i \omega t}$, such that Eq.~(\ref{TimeRadialEq}) becomes
    \begin{align}
    \left\{f(r)\frac{d}{dr}\left[f(r)\frac{d}{dr}\right]-V_{II}(r)\right\}u(r)=0, \label{RadialEq}
\end{align}
where the effective potential $V_{II}(r)$ is given by
    \begin{align}
    V_{II}(r)\equiv -\left[\omega-q\Phi(r)\right]^2 + f(r)\left[\mu^2+\frac{f'(r)}{r}+\frac{\ell(\ell+1)}{r^2}\right]. \label{EffectivePotential}
\end{align}

\section{Superradiant instability: time-domain analysis}

In this section, we perform time-domain evolutions of the scalar field perturbation. Introducing the tortoise coordinate through $dr_\ast=dr/f(r)$, the time-radial equation~(\ref{TimeRadialEq}) becomes
\begin{align}
    -\frac{\partial^2 \Psi}{\partial t^2} + \frac{\partial^2 \Psi}{\partial r_\ast^2}- 2 i q \Phi \frac{\partial \Psi}{\partial t} - V_I(r) \Psi = 0. \label{TimeRadialEq1}
\end{align}
Following Refs.~\cite{Abdalla:2010nq, Zhu:2014sya}, we solve this equation using a finite-difference scheme. With the discretization $\Psi(t,r_\ast)=\Psi(i\Delta t,j\Delta r_\ast)=\Psi_{i,j}$, $\Phi(r(r_\ast))=\Phi(j\Delta r_\ast)=\Phi_j$, and $V_I(r(r_\ast))=V_I(j\Delta r_\ast)=V_{Ij}$, a second-order accurate discretization ${\cal O}(\Delta t^2,\Delta r_\ast^2)$ yields

    \begin{align}
    - \frac{\Psi_{i+1, j} - 2 \Psi_{i, j} + \Psi_{i-1, j}}{\Delta t^2} -2 i q \Phi_j \frac{\Psi_{i+1, j}- \Psi_{i-1, j}}{2 \Delta t} + \frac{\Psi_{i, j+1} - 2 \Psi_{i, j} + \Psi_{i, j-1}}{\Delta r_\ast^2} - V_{I j} \Psi_{i,j} =0. \label{TimeRadialEq2}
\end{align}

We choose the initial perturbation to be a Gaussian wave packet,
$\Psi(t=0,r_\ast)=\exp\left[-\frac{(r_\ast-a)^2}{2b^2}\right]$,
and fix $a=10$ and $b=3$ for definiteness. To ensure numerical stability, we set $\Delta t/\Delta r_\ast=0.5$, which satisfies the von Neumann stability condition.

In the tortoise coordinate, both boundaries are located at infinity. In practice, the numerical domain must be truncated, which can introduce spurious boundary reflections. To mitigate this effect, we place the boundaries sufficiently far from the extraction point so that any reflected signal does not reach it within the simulated time interval.

\begin{figure}[!htbp]
    \centering
    \includegraphics[width=0.46\linewidth]{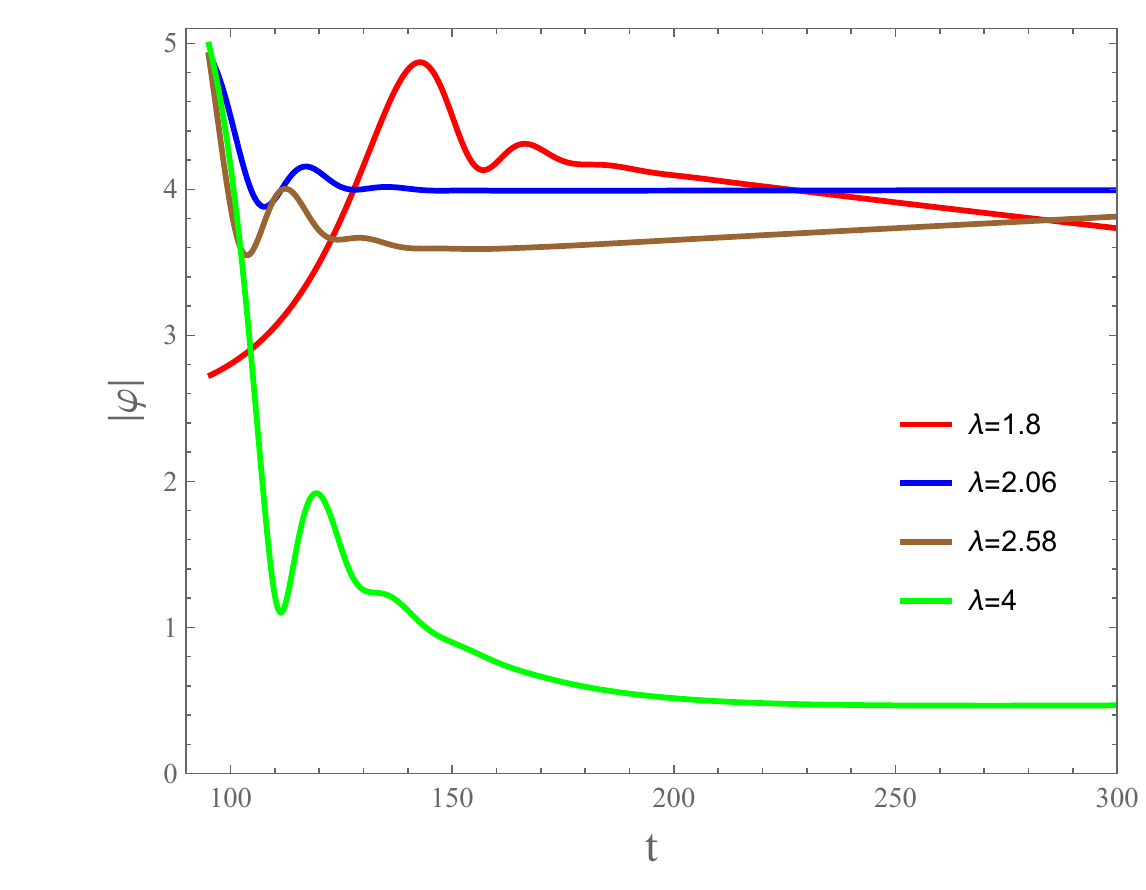}\quad
    \includegraphics[width=0.46\linewidth]{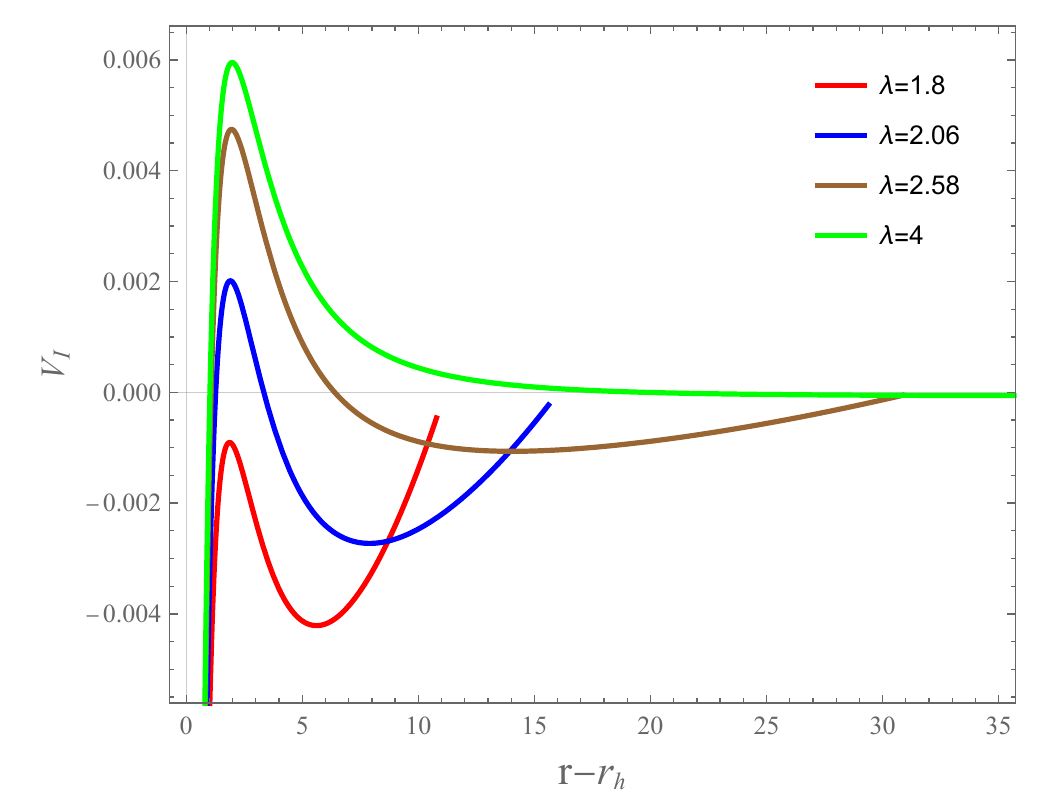}
    \caption{{\em Left}: Time evolution of the massless scalar field perturbations $\varphi \equiv \Psi/r$ for various $\Lambda = 10^{-\lambda}$. {\em Right}: Profiles of the corresponding effective potential $V_{I} (r)$. We set $Q=0.4, q=0.5$ and $\ell=0$. In the present case, instability only occurs when $\lambda \gtrsim 2.06$ and peaks at $\lambda \simeq 2.58$. All quantities are measured in units of $M$.}
    \label{lambdaTimeDomain}
\end{figure}

\begin{figure}[!htbp]
    \centering
    \includegraphics[width=0.46\linewidth]{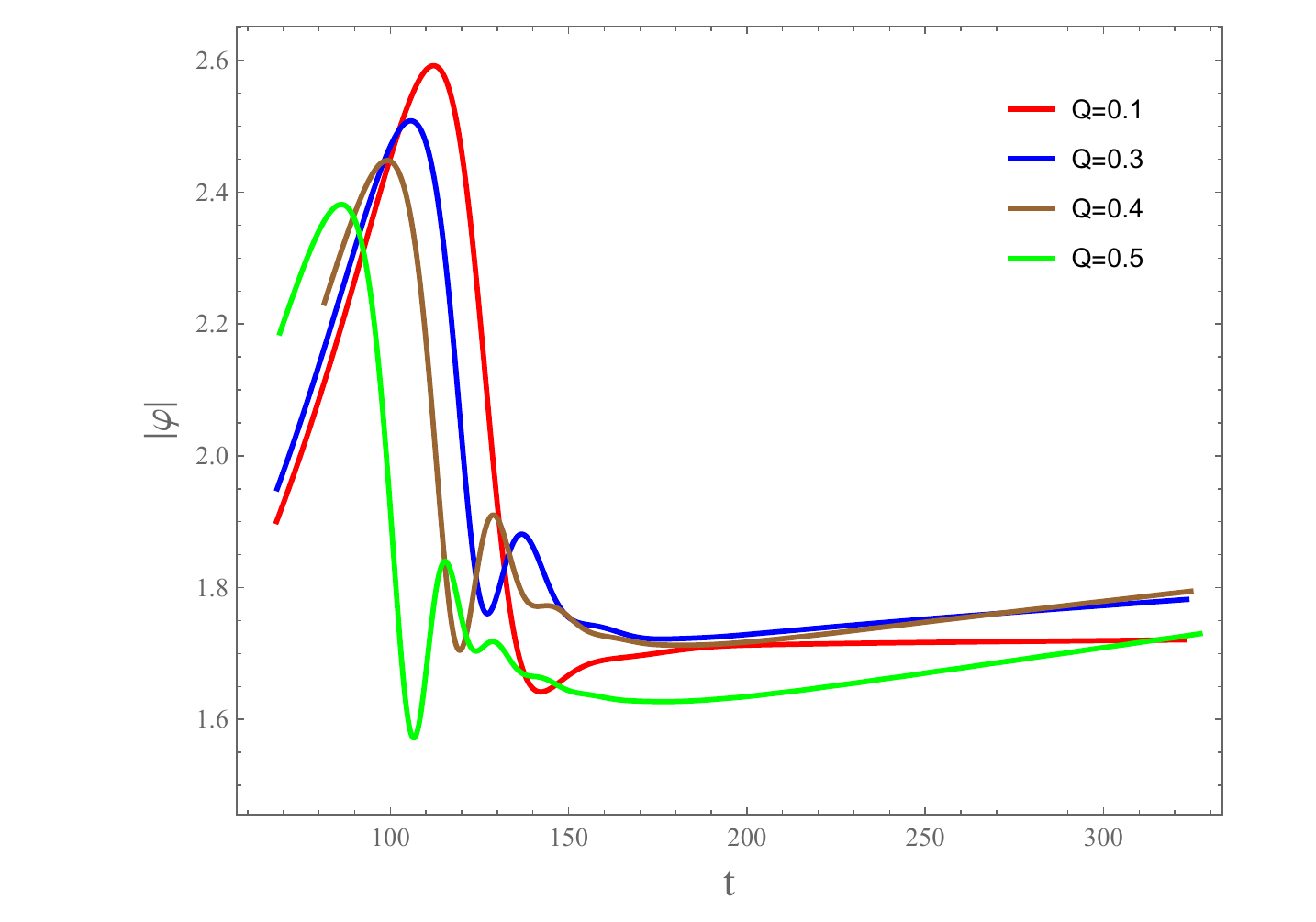}\quad
    \includegraphics[width=0.46\linewidth]{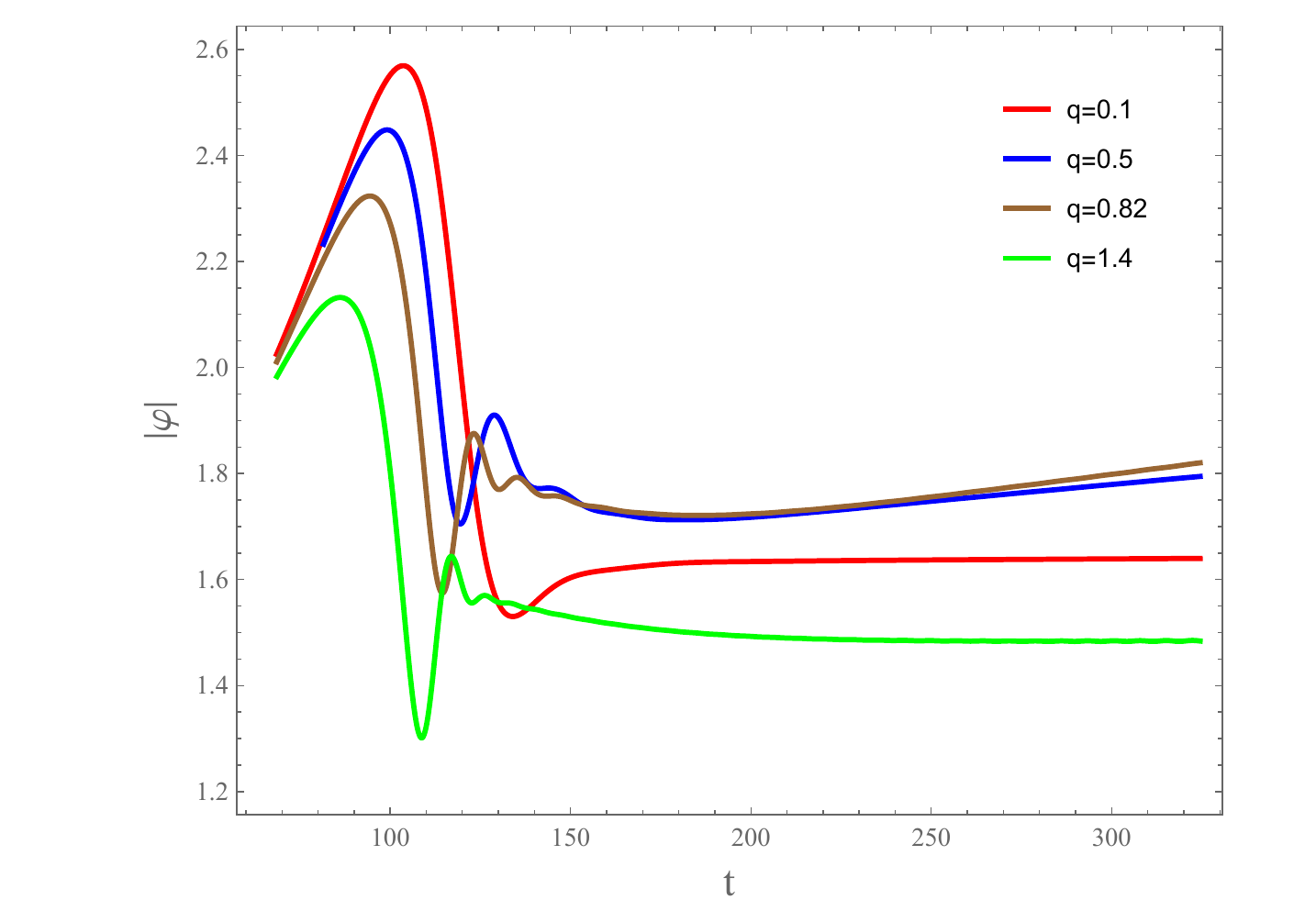}
    \caption{Time evolution of the massless scalar field perturbations $\varphi \equiv \Psi/r$ for various $Q$ and $q$. We set $\Lambda=10^{-3}$ and $\ell=0$. In the left panel, we fix $q=0.5$. In the right panel, we fix $Q=0.4$ and the divergence rate of $|\varphi|$ peaks at $q \simeq 0.82$. All quantities are measured in units of $M$.}
    \label{QqTimeDomain}
\end{figure}

The evolution is controlled by five parameters, $\{\Lambda,Q,q,\ell,\mu\}$, appearing in Eqs.~(\ref{TimeRadialEq1}) and~(\ref{TimeRadialEq2}). For simplicity, we restrict to the massless case $\mu=0$ and focus on the dependence on $\{\Lambda,Q,q,\ell\}$. We present representative examples in Figs.~\ref{lambdaTimeDomain} and~\ref{QqTimeDomain}, where $\varphi\equiv\Psi/r$. Late-time growth of the perturbation indicates an instability. Following the arguments of Ref.~\cite{Konoplya:2014lha}, we attribute this behavior to superradiance. We also note that, in the asymptotically flat limit, ABG black holes do not exhibit superradiant instability for massless charged scalar perturbations \cite{Hod:2024aen,dePaula:2024xnd,Dolan:2024qqr}.

According to the left panel of Fig.~\ref{lambdaTimeDomain}, for fixed $Q$, $q$, and $\ell=0$, the evolution is stable when $\Lambda$ is sufficiently large. Instability sets in only after $\Lambda$ drops below a critical value (corresponding to $\lambda \simeq 2.06$), which depends on the other parameters. As $\Lambda$ decreases further, the growth rate increases, reaches a maximum near $\lambda \simeq 2.58$, and then gradually decreases, approaching stability again in the limit $\Lambda \to 0$.

This behavior can be understood qualitatively from the effective potential $V_I(r)$ shown in the right panel. Superradiant instability requires a potential well that is sufficiently deep and wide to trap amplified modes. Such a well exists only for $\lambda \gtrsim 2.06$. As $\Lambda$ decreases (equivalently, as $\lambda$ increases) up to $\lambda \simeq 2.58$, the well deepens and widens, which enhances the instability. For smaller $\Lambda$, the well becomes narrower and shallower and eventually disappears as $\Lambda \to 0$, which quenches the instability.

Next, we examine the dependence on $Q$ and $q$ in Fig.~\ref{QqTimeDomain}. Here we fix $\Lambda = 10^{-3}$ and set $\ell=0$. The left panel shows that, for fixed $q$, the growth of $|\varphi|$ becomes faster as $Q$ increases. In contrast, the right panel shows a non-monotonic dependence on $q$. For fixed $Q$, the growth rate of $|\varphi|$ increases at small $q$ and decreases at larger $q$. These trends can again be interpreted qualitatively in terms of the effective potential.

\begin{figure}[!htbp]
    \centering
    \includegraphics[width=0.46\linewidth]{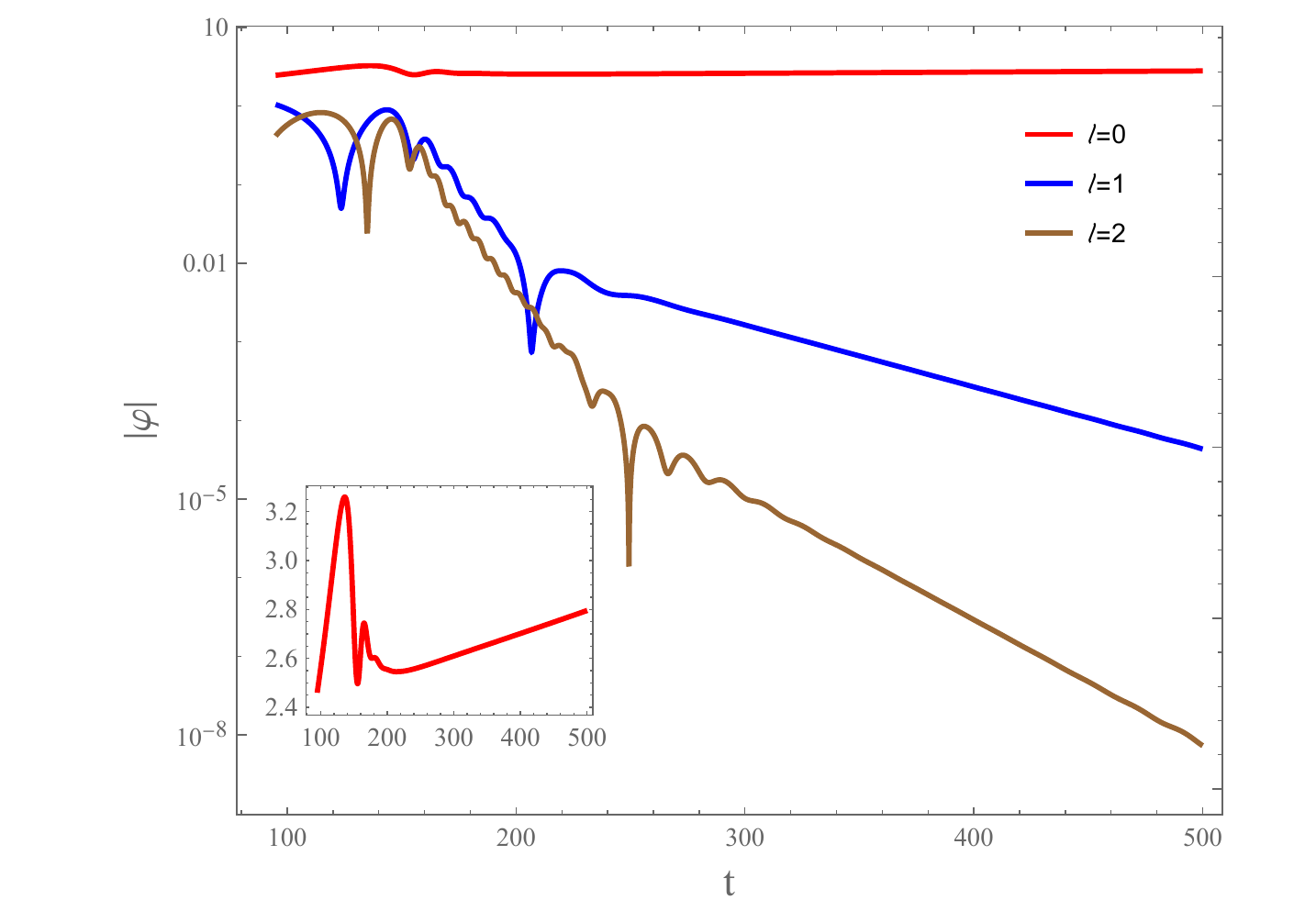}\quad
    \includegraphics[width=0.46\linewidth]{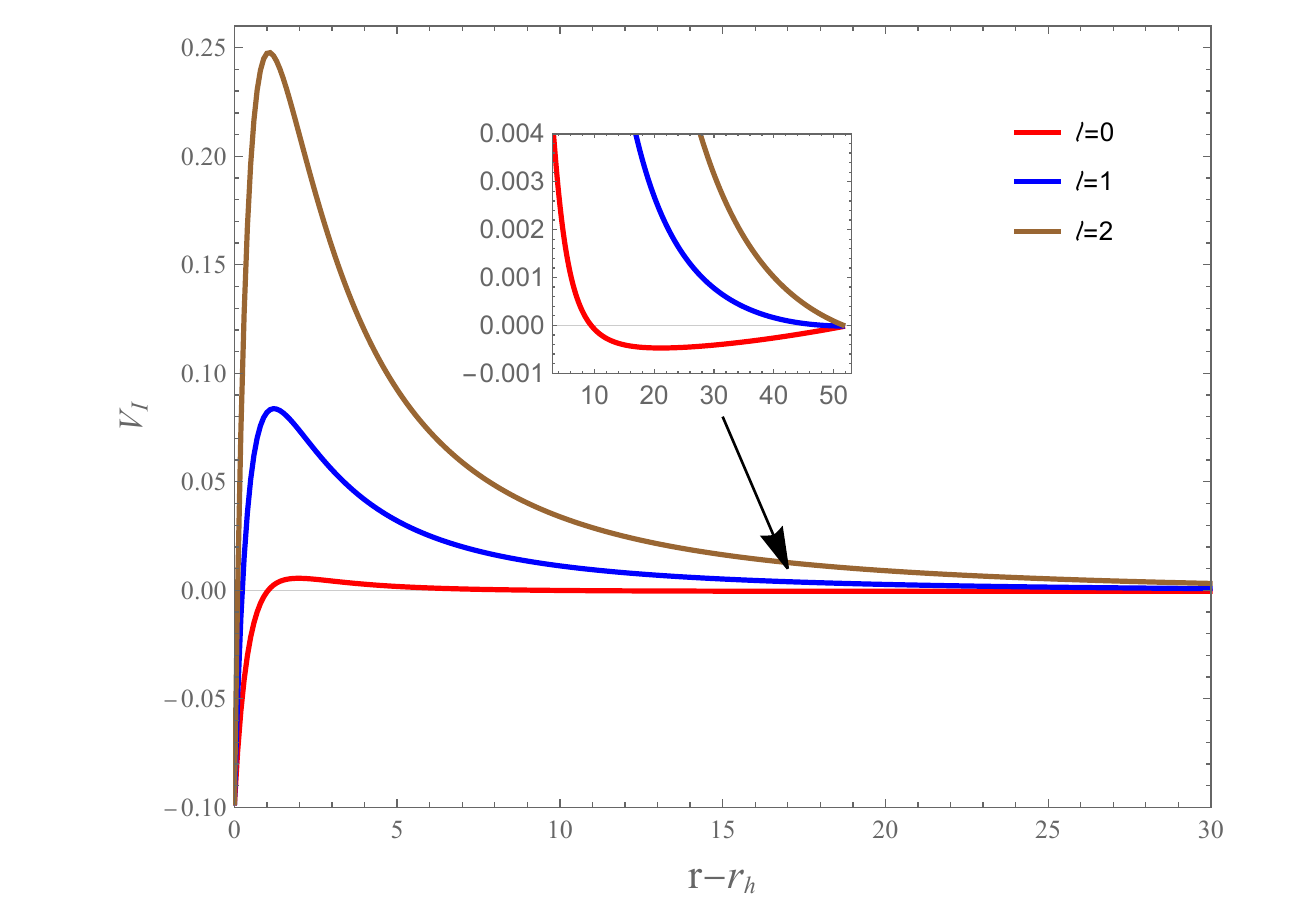}\quad
    \caption{{\em Left}: Time evolution of the massless scalar field perturbations $\varphi \equiv \Psi/r$ for various $\ell$. {\em Right}:  Profiles of the corresponding effective potential $V_{I} (r)$. We set $\Lambda=10^{-3}, Q=0.4$ and $q=0.5$. All quantities are measured in units of $M$.}
    \label{LTimeDomain}
\end{figure}
 
In the time-domain analyses above, we fixed the multipole index to $\ell=0$. We find that superradiant instability occurs only for $\ell=0$, while the evolution remains stable for all higher multipoles ($\ell>0$). The left panel of Fig.~\ref{LTimeDomain} shows representative evolutions for different $\ell$ and confirms the absence of instability when $\ell>0$. This behavior is consistent with previous results for RN-dS spacetimes \cite{Zhu:2014sya}. A qualitative explanation is provided by the effective potential $V_I(r)$ in the right panel. For $\ell>0$, the centrifugal term raises $V_I(r)$ so that no potential well develops. Without a confining well to trap superradiant modes, the amplification mechanism cannot lead to an instability. For $\ell=0$, by contrast, a well of sufficient depth and width can form and sustain the superradiant amplification cycle.

\section{Superradiant instability: frequency-domain analysis}

In this section, we study the superradiant instability in the frequency domain by solving the radial equation~(\ref{RadialEq}) with physically motivated boundary conditions. With the tortoise coordinate, Eq.~(\ref{RadialEq}) can be written in a Schr\"odinger-type form,
\begin{align}
    \left\{\frac{d^2}{dr_\ast^2} -V_{II}(r)\right\}u(r)=0, \label{RadialEq1}
\end{align} 
where the effective potential $V_{II}(r)$ has the asymptotic behavior
\begin{eqnarray}
    V_{II}(r) \rightarrow \left\{
    \begin{array}{cc}
    - (\omega -\omega_c)^2, & \quad r \rightarrow r_h,\\
    - (\omega -\omega_{cc})^2, & \quad r \rightarrow r_c,
    \end{array}
    \right.\label{EffectivepotentialAsymptoticBehavior}
\end{eqnarray}
and the two critical frequencies are defined by $\omega_c\equiv q\Phi(r_h)$ and $\omega_{cc} \equiv q\Phi(r_c)$. 
With Eqs.~(\ref{RadialEq1}) and~(\ref{EffectivepotentialAsymptoticBehavior}), we impose the boundary conditions
\begin{eqnarray}
u(r) \rightarrow \left\{
\begin{array}{cc}
e^{- i (\omega -\omega_c) r_\ast}, & \quad r \rightarrow r_h,\\
e^{+ i (\omega -\omega_{cc}) r_\ast}, & \quad r \rightarrow r_c,
\end{array}
\right.\label{BoundaryConditions}
\end{eqnarray}
which correspond to purely ingoing waves at the event horizon and purely outgoing waves at the cosmological horizon.

With the boundary conditions~(\ref{BoundaryConditions}), the mode frequency $\omega$ is determined numerically by solving the radial equation~(\ref{RadialEq}). We use the direct integration method \cite{Pani:2013pma}; see also Ref.~\cite{Zhan:2024gvi} for a detailed description. The equation depends on the parameters $\{\Lambda,Q,\mu,q,\ell\}$. As in the time-domain analysis, we set $\mu=0$. Furthermore, since our time-domain results indicate that instability arises only for $\ell=0$, we set $\ell=0$ throughout the frequency-domain analysis. Superradiance occurs when the real part of $\omega$ satisfies \cite{Zhu:2014sya, Konoplya:2014lha} (see also the review \cite{Brito:2015oca})
\begin{align}
    \omega_{cc} < \operatorname{Re} \omega < \omega_c. \label{SuperradianceCondition}
\end{align}
If the imaginary part satisfies $\operatorname{Im} \omega>0$, the corresponding mode grows exponentially in time and signals an instability.

\begin{figure}[!htbp]
\centering
\includegraphics[width=0.46\textwidth]{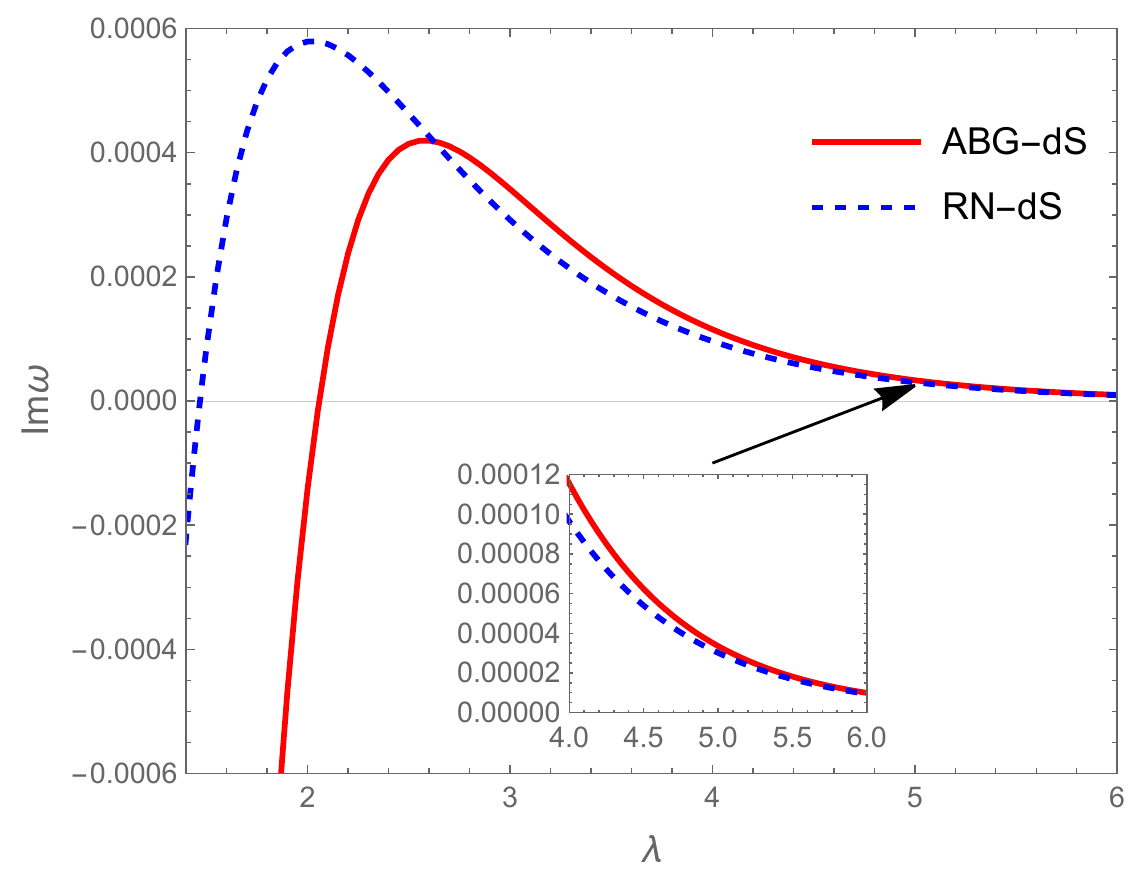}\quad
\includegraphics[width=0.46\textwidth]{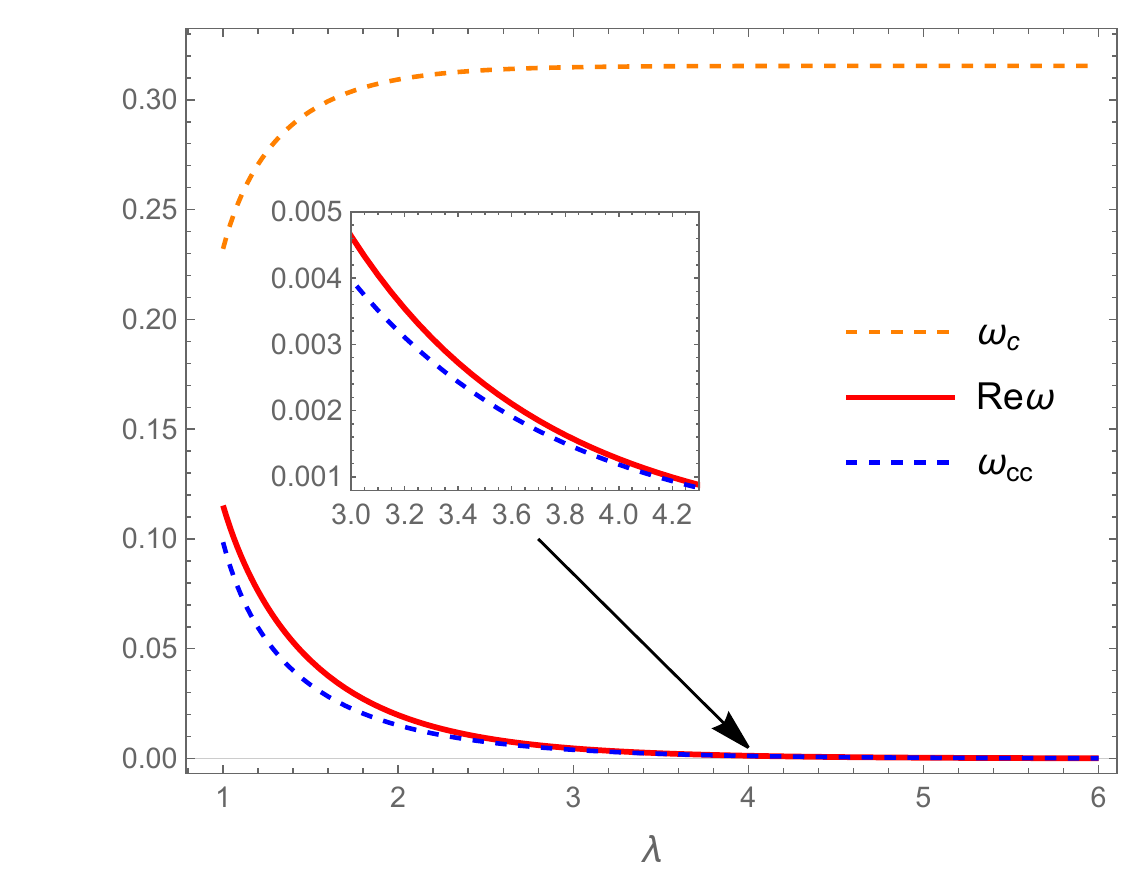}\\
\caption{Fundamental modes as a function of $\Lambda \equiv 10^{-\lambda}$. Here we set $Q=0.4, q=0.5$ and $\ell=0$. All quantities are measured in units of $M$.}
\label{omega-Lambda}
\end{figure}

\begin{figure}[!htbp]
\centering
\includegraphics[width=0.46\textwidth]{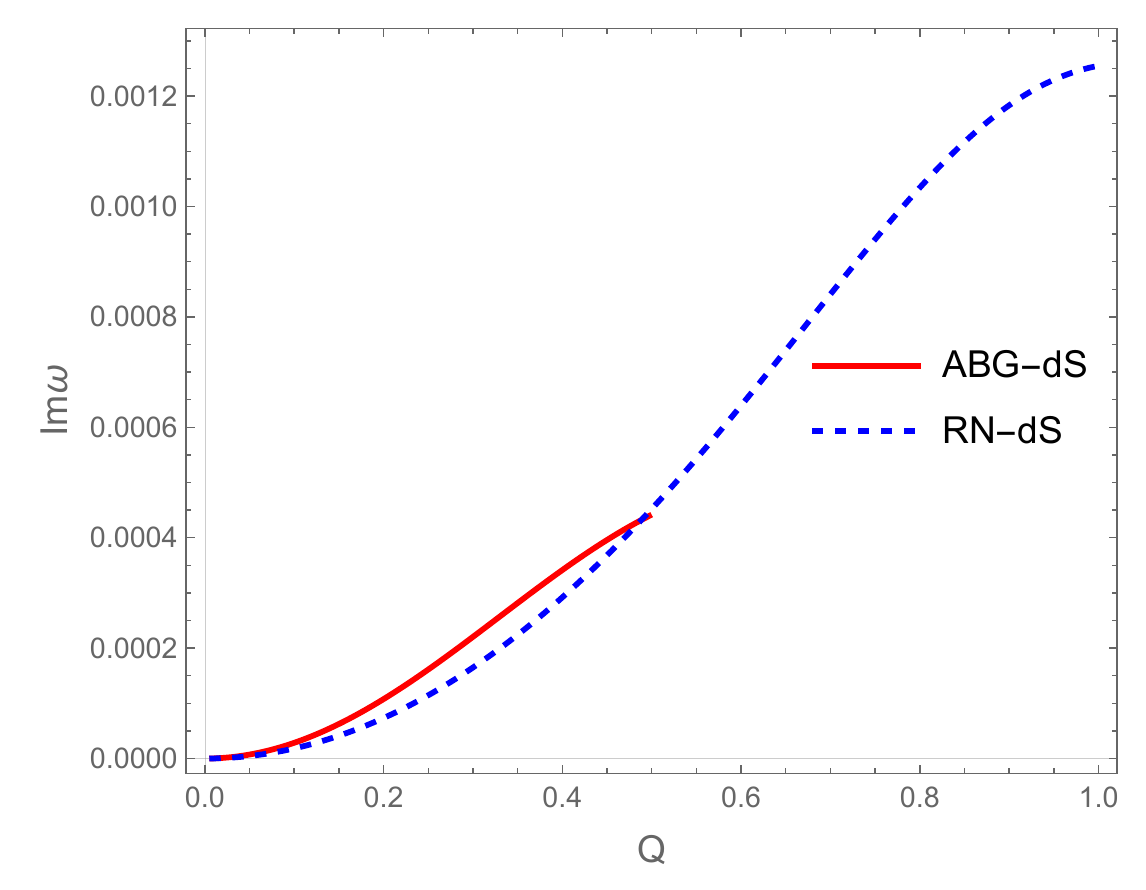}\quad
\includegraphics[width=0.46\textwidth]{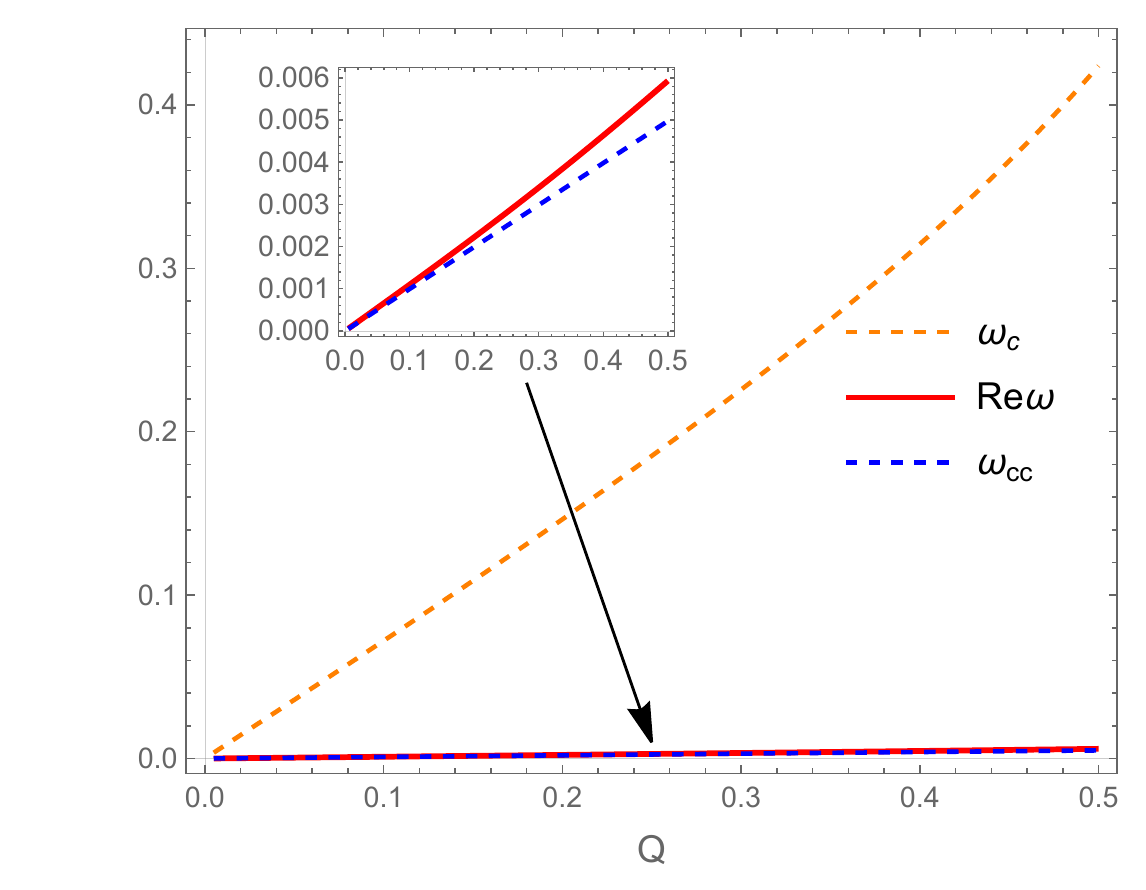}\\
\caption{Fundamental modes as a function of $Q$. Here we set $\Lambda=10^{-3}, q=0.5$ and $\ell=0$. All quantities are measured in units of $M$.}
\label{omega-Q}
\end{figure}

\begin{figure}[!htbp]
\centering
\includegraphics[width=0.46\textwidth]{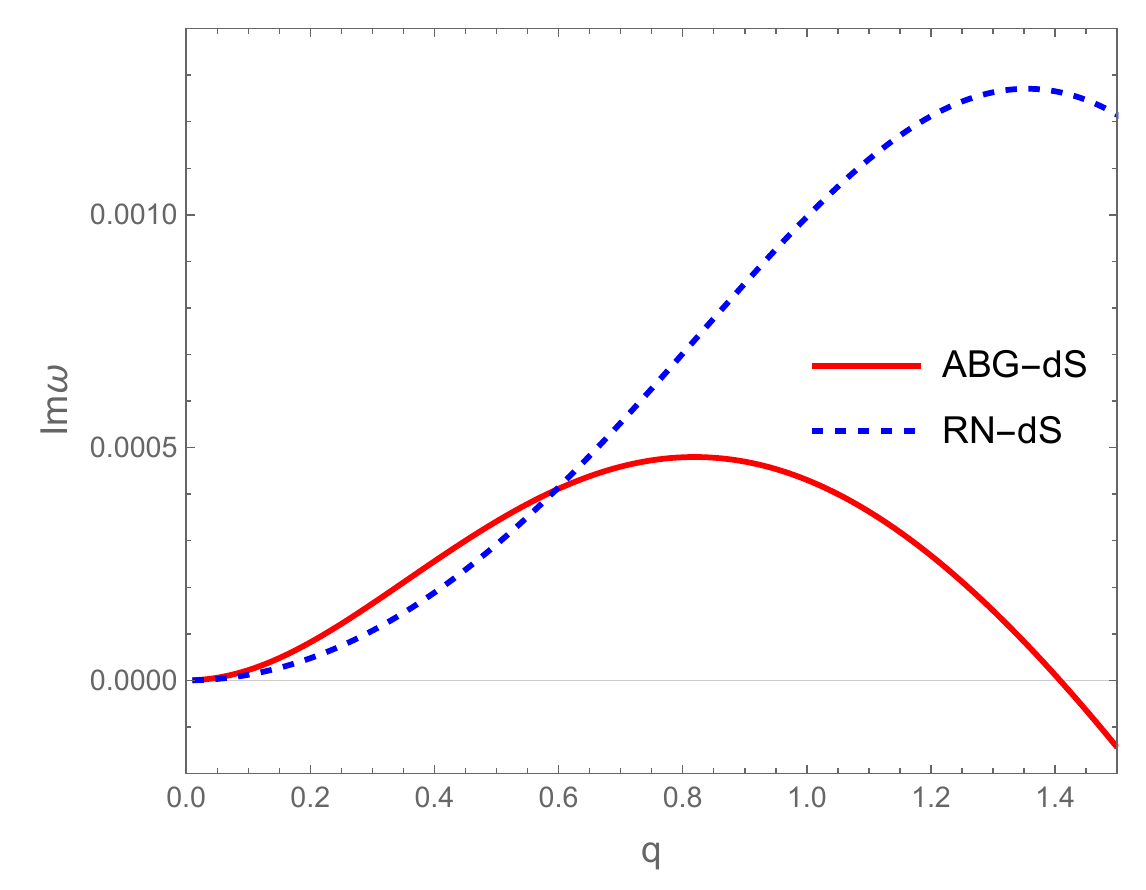}\quad
\includegraphics[width=0.46\textwidth]{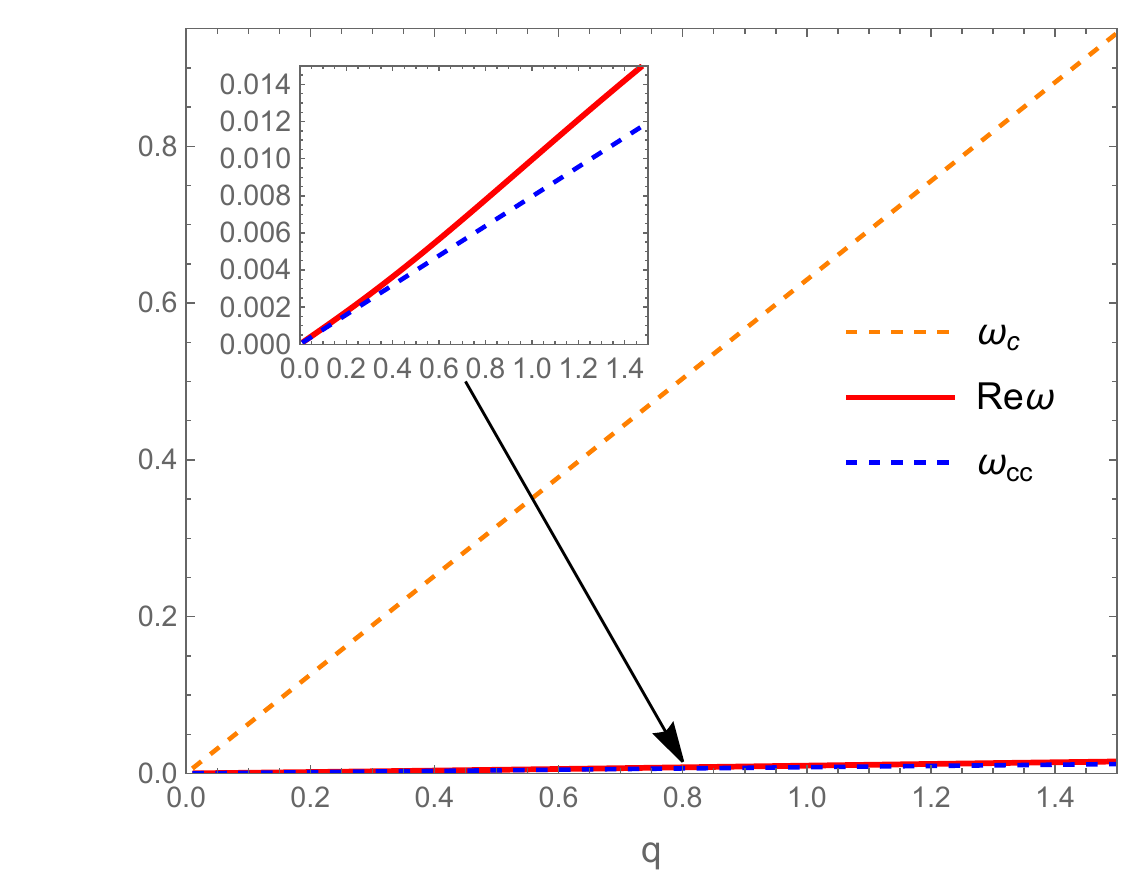}\\
\caption{Fundamental modes as a function of $q$. Here we set $\Lambda=10^{-3}, Q=0.4$ and $\ell=0$. All quantities are measured in units of $M$.}
\label{omega-q}
\end{figure}

The numerical results for $\omega$ at representative parameter values are shown in Figs.~\ref{omega-Lambda}, \ref{omega-Q}, and \ref{omega-q}. We use the same parameter choices as in the time-domain analysis to facilitate a direct comparison. For reference, we also include $\operatorname{Im}\omega$ for the RN-dS case. The left panels show that instability, namely $\operatorname{Im}\omega>0$, occurs only within specific parameter ranges. The right panels confirm that $\operatorname{Re}\omega$ satisfies the superradiance condition~\eqref{SuperradianceCondition}, which supports the superradiant origin of the instability. In addition, $\operatorname{Re}\omega$ stays close to the lower bound $\omega_{cc}$ across the parameter ranges considered.

From the left panel of Fig.~\ref{omega-Lambda}, for fixed values of the other parameters, $\operatorname{Im}\omega$ becomes positive at $\lambda \gtrsim 2.06$, peaks near $\lambda \simeq 2.58$, and gradually decays toward zero as $\Lambda \to 0$. This behavior agrees with the time-domain result shown in Fig.~\ref{lambdaTimeDomain}. In comparison, the RN-dS model becomes unstable at a larger $\Lambda$, corresponding to $\lambda \gtrsim 1.55$, reaches a higher peak near $\lambda \simeq 2.0$, and decays more steeply. Consequently, RN-dS exhibits a stronger instability than ABG-dS when $\lambda \lesssim 2.6$. For sufficiently small $\Lambda$, both models approach nearly the same small positive value of $\operatorname{Im}\omega$, which indicates that the instability disappears in the limit $\Lambda \to 0$.

From the left panel of Fig.~\ref{omega-Q}, with all other parameters fixed, $\operatorname{Im}\omega$ is close to zero at small $Q$ and increases monotonically as $Q$ increases. This indicates that a larger black hole charge leads to a stronger instability. This trend is consistent with the time-domain analysis in Fig.~\ref{QqTimeDomain}. In comparison, the RN-dS case shows a slower increase of $\operatorname{Im}\omega$, and it remains below the ABG-dS curve for $Q \lesssim 0.5$.

From the left panel of Fig.~\ref{omega-q}, with all other parameters fixed, $\operatorname{Im}\omega$ for the ABG-dS black hole is close to zero at small $q$, increases to a maximum near $q \simeq 0.82$, and then decreases, becoming negative around $q \simeq 1.4$. This indicates that the instability exists only within a finite interval of $q$, with the largest growth rate attained at intermediate values. This behavior is consistent with the time-domain results in Fig.~\ref{QqTimeDomain}. In comparison, the RN-dS case is also stable at small $q$, but $\operatorname{Im}\omega$ increases more slowly, continues to grow beyond $q \simeq 0.82$, reaches a maximum near $q \simeq 1.3$, and then decreases slightly. For $q\lesssim 0.6$, ABG-dS exhibits a stronger instability than RN-dS, whereas for $q>0.6$ the instability is weaker. The two curves diverge at large $q$ because ABG-dS becomes stable again, while RN-dS retains $\operatorname{Im}\omega>0$, corresponding to a wider instability window.

\section{Summary and discussions}

We have investigated the superradiant instability of ABG-dS black holes under massless charged scalar field perturbations. We combined time-domain evolutions with frequency-domain computations of unstable quasinormal modes.

Our results show that the instability is restricted to the spherically symmetric mode with $\ell=0$. In the asymptotically flat limit $\Lambda\to 0$, the corresponding ABG black holes are stable for massless charged scalar fields. This highlights the key role of the cosmological horizon, which provides an effective confinement mechanism that enables repeated superradiant amplification.

We also clarified how the growth rate depends on the parameters. For fixed $Q$ and $q$, instability occurs only within a finite interval of $\Lambda$, reaches a maximum at an intermediate $\Lambda$, and weakens as $\Lambda\to 0$. For fixed $\Lambda$ and $q$, the growth rate increases monotonically with the black hole charge $Q$. For fixed $\Lambda$ and $Q$, the growth rate increases with $q$ at small $q$, reaches a maximum at an intermediate value, and decreases at larger $q$.

Compared with the RN-dS case, ABG-dS black holes exhibit quantitatively different instability windows and growth rates. This difference can be attributed to the modified electrostatic potential induced by nonlinear electrodynamics.

These results contribute to the broader understanding of the stability of regular black holes in de Sitter backgrounds and may be useful for distinguishing regular solutions from their singular counterparts. It would be worthwhile to extend this work by studying the nonlinear development of the instability and assessing potential observational consequences.

\begin{acknowledgments}
	This work is supported by the National Natural Science Foundation of China (NNSFC) under Grant No 12075207.
\end{acknowledgments}

\bibliographystyle{utphys}
\bibliography{ref}

\end{document}